\documentclass[conference]{IEEEtran}
\IEEEoverridecommandlockouts
\usepackage{cite}
\usepackage{amsmath,amssymb,amsfonts}
\usepackage{algorithm}
\usepackage{algpseudocode}   
  
\usepackage{graphicx}
\usepackage{textcomp}
\usepackage{xcolor}
\usepackage{array}
\usepackage[caption=false,font=footnotesize]{subfig}
\usepackage{tabularx}        

\usepackage{tikz}
\usetikzlibrary{positioning,arrows.meta}

\def\BibTeX{{\rm B\kern-.05em{\sc i\kern-.025em b}\kern-.08em
    T\kern-.1667em\lower.7ex\hbox{E}\kern-.125emX}}
\begin{document}

\title{Hybrid Centralized–Distributed Control for Lifelong MAPF over Wireless Connections\\

\thanks{J. Cao, W. Liu, Y. Li, and B. Vucetic are with the School of Electrical and
Computer Engineering, The University of Sydney, Sydney, NSW 2006,
Australia (e-mail: jinghao.cao@sydney.edu.au; wanchun.liu@sydney.edu.au; yonghui.li@sydney.edu.au; branka.vucetic@sydney.edu.au)}
}

\author{Jinghao Cao, Wanchun Liu, Yonghui Li, Branka Vucetic}

\maketitle

\begin{abstract}
In lifelong multi-agent path finding (MAPF) with many robots, unreliable
wireless links and stochastic executions are the norm. Existing
approaches typically either rely on centralized planning under idealized
communication, or run fully distributed local controllers with fixed
communication patterns; they rarely couple communication scheduling with
policy learning, and thus struggle when bandwidth is scarce or packets
are frequently dropped. We address this joint control--communication
problem and propose a hybrid centralized--distributed scheme: a
centralized cloud policy sends small residual corrections only when
selected, while a lightweight on-board Gated recurrent unit (GRU) policy provides a safe
default fallback when wireless connection is not available.

The hybrid policy learns who/when/what to communicate through three
components: (i) a centralized policy that co-designs downlink scheduling
and residual strength on a masked belief map, focusing scarce bandwidth
on congested conflict regions; (ii) an event-aware risk module that
turns vertex/edge collisions, wall hits and queuing into dense,
interpretable learning signals; and (iii) link-level alignment based on
finite-blocklength reliability, yielding an uplink selection objective
that is monotone submodular (greedy $1-1/e$). With sufficient bandwidth
and reliable links, the hybrid controller recovers the global
coordination benefits of centralized planning via cloud overrides,
outperforming purely distributed baselines under uncertainty; under
scarce bandwidth and packet loss it retains the scalability and
robustness of distributed control through the local fallback. On
multiple MovingAI benchmarks under high congestion, our method yields
substantial gains in Total Number of Completed Tasks (TNCT) / throughput
and decision latency under constrained bandwidth, with near
map-size-invariant runtime scaling.

\end{abstract}

\begin{IEEEkeywords}
Multi-agent path finding, hybrid cloud--edge control, unreliable communication, finite blocklength, submodular optimization, lifelong MAPF
\end{IEEEkeywords}

\section{Introduction}
In multi-agent path finding (MAPF), autonomous agents must navigate efficiently and collision-free in shared spaces such as warehouses, UAV swarms, and automated guided vehicles. Despite steady progress, much of the literature still relies on idealized assumptions, with limited integration of communication failures and stochastic transition uncertainty in agent motion (the random outcome of movement commands). As a result, method applicability degrades under unstable links, congested bottlenecks, and large-team coordination. Moreover, mainstream approaches often sit at the extremes of fully centralized or fully decentralized: centralized schemes leverage global information but struggle to scale, whereas decentralized policies react quickly yet can lack global coordination and solution quality. Recent learning-based advances—distributed policy learning, lifelong MAPF, and selective communication via messages/attention—improve scalability and cooperation \cite{sartoretti2019primal,9424371,damani2021primal,lin2023sacha}, but a cloud–edge hybrid formulation that jointly aligns communication constraints, execution risks, and policy optimization remains underexplored.

To close this gap, we propose a Collaborative Distributed-Centralised (CDC) framework with two complementary modules trained in sequence: a lightweight on-board GRU policy for standalone execution, and a cloud residual corrector that refines actions whenever communication is available. The modules use interface-aligned, heterogeneous, multi-source inputs; when connected, the cloud injects a residual into the on-board logits, while under outages the edge provides a robust fallback. In addition, we introduce a belief-map representation that encodes environment structure and neighbors’ short-horizon intentions—derived from uploaded hidden states and cloud rollouts—under communication and execution uncertainty, thereby facilitating cooperation among agents and strengthening fallback decisions when direct information is unavailable. Grounded in experience with practical wireless channels for remote control \cite{10702475}, our training and evaluation incorporate finite-blocklength success probabilities \cite{9134368} and event-level risk modeling, yielding significant gains over purely on-board and message-free distributed baselines under constrained bandwidth and high congestion. This hybrid design approaches centralized behavior under reliable links and preserves distributed scalability and robustness otherwise, at the price of an extra cloud–edge coordination layer and its associated overhead.

\section{Related Work}
In MAPF, methods cluster at two extremes: fully centralized (e.g., ODrM and heuristic/anytime variants) or fully decentralized (e.g., PRIMAL with message passing, priorities, or auctions); truly hybrid distributed--centralized schemes are rare and usually limited to short-range joint replanning near detected conflicts or hierarchical task--path pipelines without a systematic communication--control treatment~\cite{ferner2013odrm,huang2022anytime,sartoretti2019primal,li2021lifelong,skrynnik2024learn,ma2021distributed,li2022multi}.
Outside MAPF, selective-communication Multi-agent reinforcement learning (MARL) under bandwidth limits learns who/when/what to transmit via gating, scheduling, variance/temporal control, and targeted messaging~\cite{singh2018learning,kim2019learning,zhang2019efficient,zhang2020succinct,wang2019learning,ding2020learning}; these approaches are typically Centralized training for decentralized execution (CTDE) and lack a run-time centralized residual that collaborates with local policies.
A parallel robotics/control strand studies cloud--edge hybrids—migrating robot--cloud ``skills'' using latency/resource monitors~\cite{muratore2023xbot2d}, delegating global (re)planning/coordination to a cloud manager with on-robot execution~\cite{zagradjanin2019cloud}, and offloading planning/localization to nearby infrastructure~\cite{spatharakis2020switching}—with run-time CDC collaboration shown in wireless control (CDC-DRL)~\cite{wang2024wireless}. Yet these do not target MAPF's vertex/edge conflict and queue structures nor finite-blocklength link reliability; they emphasize \emph{where to compute} over action-level conflict resolution with execution-time centralized--distributed residual refinement, and seldom demonstrate scalable, real-time, sustained decongestion under high congestion among agents and unreliable links.

% We therefore propose a MAPF-specific hybrid: a robust local policy refined by a centralized residual module, with joint optimization of recipients, timing, and content under bandwidth and packet-loss models, targeting proactive relief of future conflicts and queues rather than mere traffic reduction.

On uncertainty in communication and action timing and the resulting asynchronous execution, several studies inject stochastic delays and failures at the model level. For example, MAPF-DP explicitly models probabilistic transmission and action delays and compares minimal-communication policies with fully synchronized ones \cite{ma2017multi}.
ACE introduces random action delays during training to mitigate asynchronous execution \cite{yu2023asynchronous}. Yet, much of the literature confines randomness to data sampling (random maps/starts/goals) \cite{sartoretti2019primal,skrynnik2024learn,ma2021distributed,li2022multi,li2021lifelong}, without systematically feeding communication constraints (bandwidth, loss) and execution risks (wall hits, queueing) into the policy objective, nor evaluating with \text{lifelong throughput} metrics such as TNCT.

These gaps motivate a hybrid communication–control design that learns communication and action decisions together. In our framework, the cloud decides which agents receive downlink updates, when these updates are sent, and how to adjust the on-board action through a residual term. This residual action is a small correction added to the action proposed by the on-board policy. On the communication side, we impose an explicit bandwidth budget and use a finite-blocklength physical-layer model, which is more appropriate for short control packets than an ideal infinite-blocklength capacity model. This model maps SNR and code rate to a packet-success probability for each control message. On the task side, we convert event-level risks—vertex and edge conflicts, wall hits, and queuing delays—into shaped rewards that provide clear and interpretable credit signals for policy learning.

\section{Problem Formulation}
% \subsection{MAPF statement}
% We consider MAPF on an undirected graph \(G=(V,E)\) with \(n\) agents \(i\in\{1,\dots,n\}\). Each agent starts from \(s_i\in V\) and targets a unique goal \(g_i\in V\). Time is discrete \(t=0,1,\dots\); at each step, an agent either moves to an adjacent vertex or stays. Vertex, edge, and swap conflicts are prohibited \cite{sharon2015conflict}. In our execution rule, any move that would cause such a conflict \text{fails}, and all involved agents remain at their current vertices. We distinguish \text{one-shot} MAPF (optimize a single assignment’s makespan/total path) from \text{lifelong} MAPF (upon reaching a goal, an agent immediately receives a new task), where evaluation focuses on total completions/throughput, equivalently TNCT, over a window \cite{sartoretti2019primal,li2021lifelong}. We focus on lifelong MAPF, which better reflects real deployments, and our simulations likewise concentrate on the lifelong settings.

% \subsection{Experimental Environment}
% Similar to standard MAPF problem setups, we consider a 2D grid map with discrete state-transition function. All the agents and obstacles will occupy an entire vertex, and all types of conflicts are strictly prohibited: no two different entities may occupy the same vertex at any time. The grid map itself could be describe by a $m_x\times m_y$ matrix where 0 represents an empty cell, 1 represent an occupied cell. An example obstacle layout is shown in Fig.\ref{fig:mapfstatement:map}.
\subsection{MAPF Problem Statement}
\label{MAPF Problem Statement}
We model a fleet of autonomous mobile robots navigating in a shared indoor workspace such as a warehouse or factory floor. The physical free space (corridors between shelves, loading docks, intersections) is discretized into an undirected graph $G=(V,E)$, where each vertex $v\in V$ represents a collision-free stopping location for a robot and each edge $(u,v)\in E$ represents a feasible motion corridor that a robot can traverse within one control cycle (see Fig.~\ref{fig:BD} for an illustration). Each robot carries its own local controller but shares the same discrete motion layer described below.

Formally, we consider MAPF on an undirected graph $G$ with $n$ agents $i\in\{1,\dots,n\}$. The environment is a 4-neighbor grid; agents act in $\mathcal{A}=\{\text{stay},\text{up},\text{down},\text{left},\text{right}\}$ with no diagonal moves, corresponding to either remaining at the current cell or moving to one of its four adjacent cells in the physical space. Each agent starts at $s_i\in V$ and targets a unique goal $g_i\in V$. Time is discrete $t=0,1,\dots$.

We prohibit vertex, edge, and swap conflicts \cite{sharon2015conflict} as a high-level abstraction of collision avoidance and simple traffic rules. Concretely, edge conflicts (two agents attempting to traverse the same edge in opposite directions) and swap conflicts (two agents attempting to exchange vertices) cause all involved moves to fail so those agents remain in place, whereas a vertex conflict (two or more agents attempting to enter the same vertex) is resolved by uniformly random tie-breaking that allows exactly one contender to enter while the others stay. We distinguish one-shot MAPF—optimizing a single assignment’s makespan or total path—from lifelong MAPF, where upon reaching a goal an agent immediately receives a new task and evaluation focuses on throughput (TNCT) over a time window \cite{sartoretti2019primal,li2021lifelong}. We adopt the lifelong setting as it better reflects continuous operation in real deployments and use it throughout our simulations.

\begin{figure}[!t]
  \centering
  \subfloat[Map with obstacles\label{fig:mapfstatement:map}]{
    \includegraphics[width=0.45\linewidth]{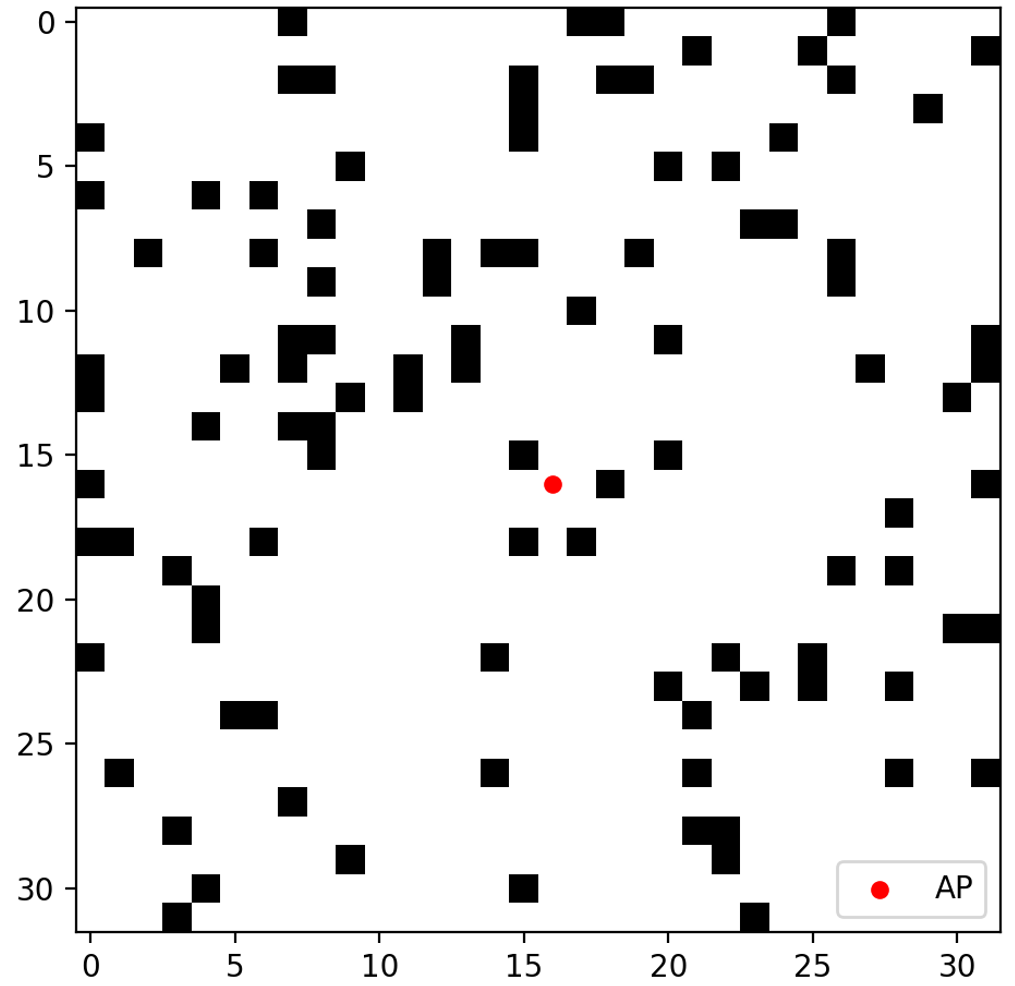}}
  \hfill
  \subfloat[Sample Radio Map\label{fig:mapfstatement:snr}]{
    \includegraphics[width=0.45\linewidth]{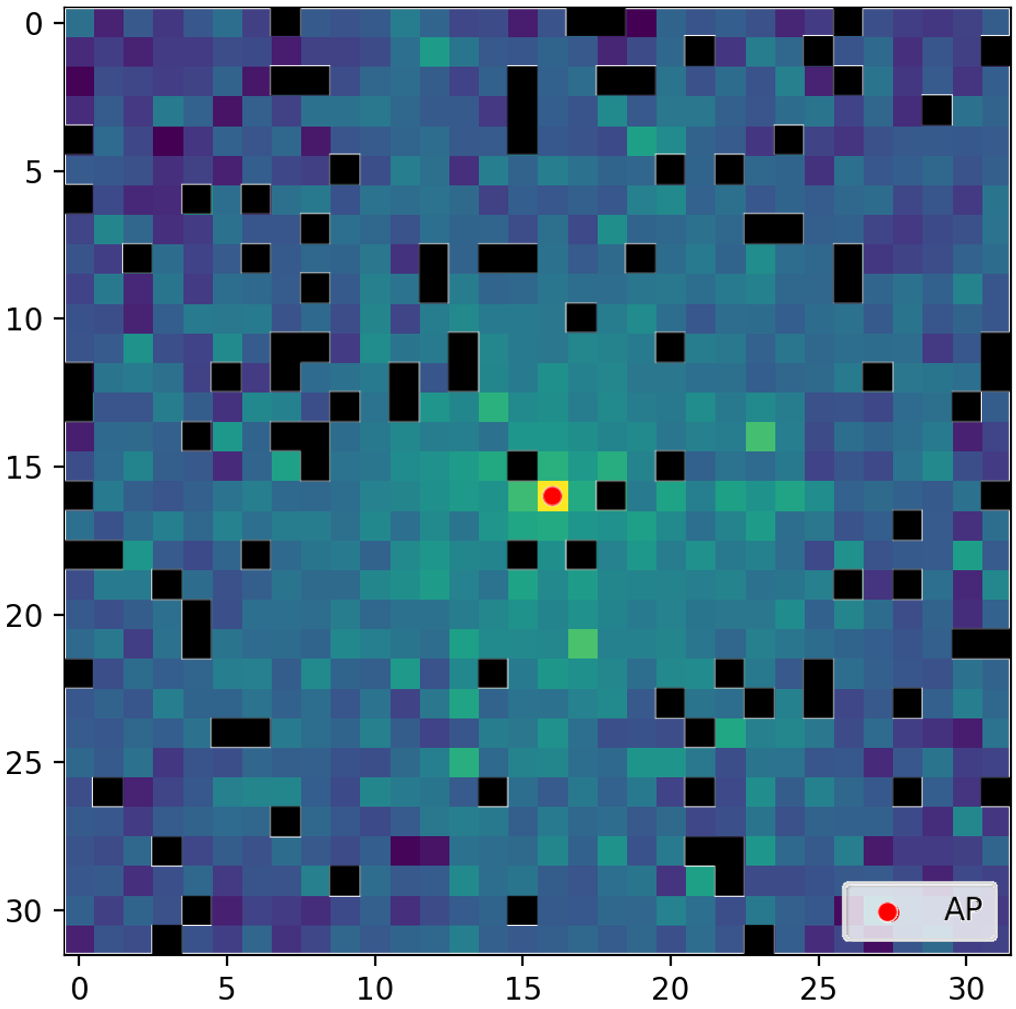}}
  \caption{Sample Obstacle map and its Radio Map.}
  \label{fig:mapfstatement}
\end{figure}

\begin{figure*}[!t]
    \centering
    \includegraphics[width=1\linewidth]{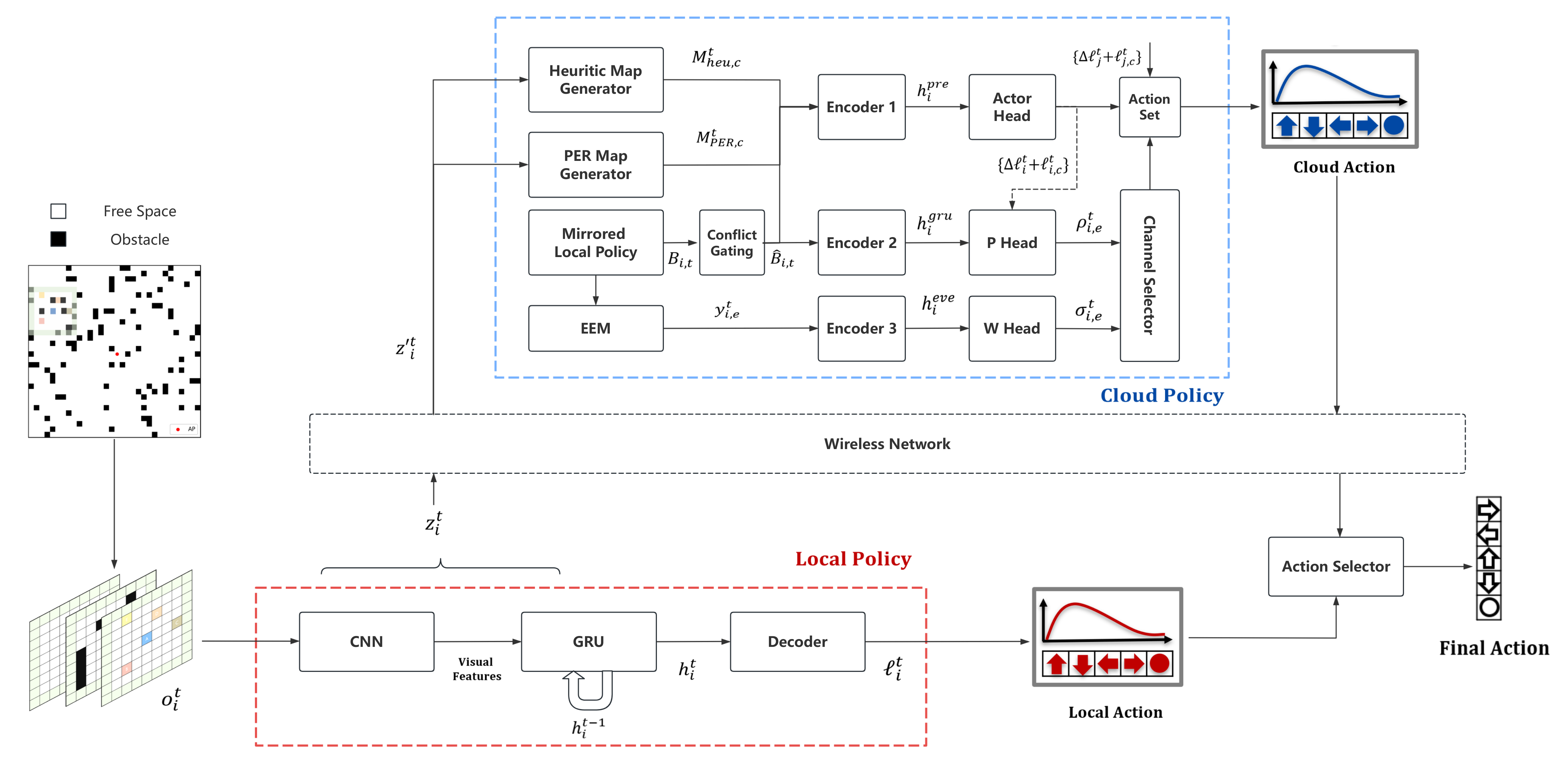}
    \caption{\textbf{Hybrid cloud--edge architecture.}
    The local GRU policy (red) generates masked action logits from $3{\times}W_{\text{FOV}}{\times}H_{\text{FOV}}$ observations and guarantees autonomy during dropouts.
    The cloud (blue) builds a belief map, predicts event risks with EEM, and uses a shared encoder plus a downlink allocator (W/P expert heads with gating) to compute residual corrections that are injected into on-board logits when the downlink is available, producing the final policy.}
    \label{fig:BD}
\end{figure*}

% \subsection{Experimental Environment}
We instantiate $G$ as a 2D grid and encode each map as an $m_x\times m_y$ binary matrix ($0$=free, $1$=obstacle). Agents or obstacles occupy single vertices and single-occupancy is enforced. All simulations use the 4-neighbor adjacency, action set, and conflict-resolution rule specified previously. A sample obstacle layout appears in Fig.~\ref{fig:mapfstatement:map}.

% \subsection{Uncertainties}
% In addition to classical MAPF problem settings, we introduced two key sources of uncertainty (execution and communication) into consideration.

% \subsection{Execution model}
We model stochastic execution by a parameterized transition kernel. Let $s_i^t \in \mathcal{S}$ be the state of agent $i$ at time $t$, and let $a_i^t \in \mathcal{A}$ denote its \emph{intended} action. In a deterministic environment, the next state is $s_i^{t+1}=f(s_i^t,a_i^t)$. Under execution uncertainty, the next state is drawn from a stochastic kernel $\mathcal{K}_\varepsilon$ that redistributes a small probability mass from the intended forward cell to stay/side/back outcomes:
\[
\varepsilon \triangleq (\varepsilon_{\text{stay}},\,\varepsilon_{\text{side}},\,\varepsilon_{\text{back}})\in[0,1]^3,\quad
\varepsilon_{\text{stay}}+\varepsilon_{\text{side}}+\varepsilon_{\text{back}}\le 1.
\]
Denote by $\mathrm{fwd}(a_i^t)$ the forward cell of $a_i^t$, by $\mathrm{side}(a_i^t)$ the two orthogonal cells, and by $\mathrm{back}(a_i^t)$ the opposite cell. Then
\begin{align*}
\Pr\!\big(s_i^{t+1}=\mathrm{fwd}(a_i^t)\big) &= 1-\varepsilon_{\text{stay}}-\varepsilon_{\text{side}}-\varepsilon_{\text{back}},\\
\Pr\!\big(s_i^{t+1}=s_i^t\big) &= \varepsilon_{\text{stay}},\\
\Pr\!\big(s_i^{t+1}\in \mathrm{side}(a_i^t)\big) &= \varepsilon_{\text{side}},\\
\Pr\!\big(s_i^{t+1}=\mathrm{back}(a_i^t)\big) &= \varepsilon_{\text{back}}.
\end{align*}

If a sampled destination is a wall or outside the grid, we apply a
bounce-to-stay rule that reassigns its probability mass to the
current state $s_i^t$ .

% We instantiate the abstract kernel with $(\varepsilon_{\text{stay}},\varepsilon_{\text{side}},\varepsilon_{\text{back}})=(0.05,0.05,0)$ unless otherwise noted, i.e., intended forward succeeds with probability $0.90$.

\subsection{Communication model}

% \noindent
\textbf{Channel Model.}
We consider a time-slotted wireless control loop with an uplink (UL, agent to server) subframe followed by a downlink (DL, server to agent) subframe. We operate in a mid-band, single-carrier setting with center frequency $f_0$ (sub-6\,GHz). The total bandwidth $B_{\mathrm{tot}}$ is partitioned into $M=\lfloor B_{\mathrm{tot}}/B_{\mathrm{RB}}\rfloor$ orthogonal time--frequency resource blocks (RBs) of bandwidth $B_{\mathrm{RB}}$. Per control step we schedule at most $C_{\mathrm{UL}}$ and $C_{\mathrm{DL}}$ concurrent RBs in UL and DL, respectively.

The radio environment is modeled as an SNR field over the grid. We assume a
single access point (AP) at a fixed location $x_{\mathrm{AP}}$.
For an agent located at position $x$, the Tx–Rx distance is
\[
  d(x) = \|x - x_{\mathrm{AP}}\|_2 \quad [\mathrm{m}].
\]
The corresponding large–scale path loss (in dB) is described by standard LoS/NLoS
models parameterized by the distance $d$ and a carrier frequency $f_0$ (in Hz):
\begin{align*}
L_{\mathrm{LOS}}(d,f_0) &= D_{\mathrm{LOS}}\log_{10}(d)+B_{\mathrm{LOS}}+F_{\mathrm{LOS}}\log_{10}(f_0),\\
L_{\mathrm{NLOS}}(d,f_0) &= D_{\mathrm{NLOS}}\log_{10}(d)+B_{\mathrm{NLOS}}+F_{\mathrm{NLOS}}\log_{10}(f_0),
\end{align*}
so higher carrier frequencies incur a larger path loss via the
$\log_{10}(f_0)$ term. For a given link state (LoS or NLoS), the average
received power on one resource block (RB) at distance $d$ is
\[
  P_{\mathrm{rx}}(d,f_0) = P_{\mathrm{tx}} - L(d,f_0) \quad [\mathrm{dBm}],
\]
where $P_{\mathrm{tx}}$ is the transmit power (per RB, in dBm) and
$L(d,f_0)$ denotes either $L_{\mathrm{LOS}}$ or $L_{\mathrm{NLOS}}$.
A sample SNR field obtained from this model is shown in
Fig.~\ref{fig:mapfstatement:map} and Fig.~\ref{fig:mapfstatement:snr}.

The per–RB SNR is then obtained from a standard link budget.
The thermal–noise power spectral density at the receiver input is
\[
  N_0 = -174~\mathrm{dBm/Hz},
\]
and the noise power on one RB of bandwidth $B_{\mathrm{RB}}$ (in Hz) with
receiver noise figure $\mathrm{NF}$ (in dB) is
\[
  P_{\mathrm{noise}} = N_0 + 10\log_{10}(B_{\mathrm{RB}}) + \mathrm{NF}
  \quad [\mathrm{dBm}].
\]
The instantaneous per–RB SNR is defined as
\[
  \gamma = \frac{P_{\mathrm{rx}}}{P_{\mathrm{noise}}}
\]
(in linear scale; we use the same symbol $\gamma$ when expressed in dB
with the usual $10\log_{10}(\cdot)$ mapping).
Here, large–scale effects are captured by $L(d,f_0)$, while
log–normal shadowing and small–scale fading are absorbed into $\gamma$
as random fluctuations around the large–scale SNR.

Time is discretized into control steps of duration $T_{\mathrm{pkt}}$ seconds.
At each step, the system is allowed to transmit one UL and one DL packet for
the scheduled agents. Over one RB, the finite–blocklength codeword
occupies
\[
  n = \eta\,B_{\mathrm{RB}} T_{\mathrm{pkt}}
\]
complex channel uses, where $0 < \eta \le 1$ accounts for overhead
(cyclic prefix, pilots, control signaling, etc.). The per–RB SNRs
$\gamma^{\mathrm{UL}}$ and $\gamma^{\mathrm{DL}}$ for the uplink
and downlink are obtained from the above link budget using the
corresponding transmit powers $P_{\mathrm{tx}}^{\mathrm{UL}}$ and
$P_{\mathrm{tx}}^{\mathrm{DL}}$.

In the UL, agent $i$ transmits a packet consists of current states $\mathbf{u}_i^t$
at step $t$; in the DL, the server returns a centralized refinement
vector $\mathbf{r}_i^t$ of size $b_{\mathrm{DL}}$ bits.
We allocate (i) which agents obtain UL/DL service and (ii) their RBs.
The transmit powers $P_{\mathrm{tx}}^{\mathrm{UL/DL}}$ are fixed,
while the coding rates $R_{\mathrm{UL/DL}}$ are selected from an MCS
table to meet a target reliability.

Packet decoding is modeled using the finite–blocklength normal
approximation for quasi–static AWGN channels~\cite{11016913,9134368}:
\begin{equation}
  p_{\mathrm{loss}}(\gamma;R,n)
  \approx
  Q\!\left(\frac{C(\gamma)-R}{\sqrt{V(\gamma)/n}}\right),
\end{equation}
where $C(\gamma)$ and $V(\gamma)$ denote the Shannon capacity and
channel dispersion, respectively, for SNR $\gamma$.
The UL and DL success probabilities are then
\begin{equation}
  p_{\mathrm{succ}}^{\mathrm{UL/DL}}
  = 1 - p_{\mathrm{loss}}\big(\gamma^{\mathrm{UL/DL}}; R_{\mathrm{UL/DL}}, n\big).
\end{equation}

% \noindent
\textbf{Uplink and Downlink.}
Instead of maps, the UL carries a list of vectors:
(i) the sender's global grid location (2-D),
(ii) locations of other agents (each 2-D),
and (iii) agents' recurrent hidden states (dimension $d_h=128$).
The list is variable-length: the sender includes itself and all other agents within its field of view (FOV). Before UL, we assume a short-range, lossless intra-team broadcast once per step so that agents
exchange hidden states; the selected UL sender then packages the collected measurements within FOV.
This assumption only concerns assembling the UL message bit-length; wide-area UL/DL still follows the
finite-blocklength reliability model already defined. As the FOV peer count increases, the UL message bit-length will grow.

% As the FOV peer count increases, the UL message bit-length will grow. With fixed RBs $m_i^t$
% and blocklength $n$, this pushes a higher $R_{\mathrm{UL}}$ so
% $\mathrm{PER}_i^t$ increases; keeping $R_{\mathrm{UL}}$ instead requires more
% RBs or packets, consuming the UL quota $C_{\mathrm{UL}}$ and adding delay.

For downlink, the server unicasts to the selected agent $i$ a bundle:
the cloud action for $i$ plus actions of other agents within $i$'s FOV.
Similar to Uplink model, the bundle size grows with the FOV peer count. So the same feature will also apply on downlink part of the model.

Finally, we couple communication with control: UL success governs the freshness and fidelity of centralized inference, while DL success determines whether residual actions arrive in time to proactively relieve future conflicts and queuing. Thus, communication success becomes a first-class decision variable aligned with MAPF risk mitigation.

\section{Proposed Method}

\subsection{Optimisation Objective}

We consider lifelong MAPF under action execution and communication uncertainties, where each agent repeatedly receives a new start--goal (start position and goal position) pair after reaching the current goal and the process continues indefinitely. The primary evaluation metric is the total number of completed tasks within a time window (TNCT), which is closely related to throughput~\cite{li2021lifelong,skrynnik2024learn}. Directly optimizing this long-horizon objective is difficult due to combinatorial constraints and sparse reward signals. Consequently, we maximize the expected discounted sum of per-step rewards, where each reward combines task-progress terms with explicit penalties for unsafe events (e.g., vertex/edge conflicts, wall hits, excessive queuing) and for communication usage, effectively encoding safety and communication costs through reward shaping.

We adopt a hybrid policy with two components. The local policy $\pi_{\text{local}}: \mathcal{H} \rightarrow \Delta(\mathcal{A})$ runs on board each agent and maps a sliding-window history $h_t \in \mathcal{H}$ to a masked distribution over actions; this local controller is always available and provides the base action proposal. When the downlink (DL) connection to the cloud is available, a cloud policy $\pi_{\text{cloud}}$ processes the uploaded information (and optionally the local logits) and returns a residual correction that refines the local distribution, yielding the final cloud-refined action distribution. When DL is unavailable, the agent simply samples and executes an action from the local policy without cloud refinement.

Formally, we learn the hybrid policy $\Pi=\{\pi_{\text{cloud}},\,\pi_{\text{local}}\}$ to maximize the expected discounted return
\[
\max_{\Pi}\ \mathbb{E}_{\pi_{\mathrm{exec}}}\!\left[\sum_{t=0}^{\infty}\beta^{t}\, r(s^{t},a^{t})\right],
\qquad 0<\beta<1,
\]
where the environment reward at time $t$ sums per-agent terms,
\[
r(s^{t},a^{t})=\sum_{i=1}^{N} r_i^t,
\]
and $N$ is the number of agents. We train both the on-board policy and the cloud residual corrector under the same per-step reward used in lifelong MAPF. Table~\ref{tab:reward-return} specifies all components and default weights. Distance shaping uses the Manhattan metric on a 4-neighbor grid:
\[
\Delta d_i^t = d_{\text{man}}(x_i^t,g_i)-d_{\text{man}}(x_i^{t+1},g_i).
\]

Here $\Delta d_t^i = d_{\mathrm{man}}(x_t^i, g_i) - d_{\mathrm{man}}(x_{t+1}^i, g_i)$ measures the one-step change in Manhattan distance from agent $i$ to its goal; a positive value means that the agent moved closer to $g_i$, while a negative value means it moved further away. 

The communication term $\tilde c_t^i \in \{0,1\}$ is defined as
\[
\tilde c_t^i =
\begin{cases}
1, & \text{if UL--DL exchange succeeded at step $t$,}\\
0, & \text{otherwise,}
\end{cases}
\]
so it indicates whether end-to-end communication is available in that control cycle.

% ; when a link is scheduled we use $\tilde c_i^t = p^{\mathrm{UL}}_{\mathrm{succ}} p^{\mathrm{DL}}_{\mathrm{succ}}$, and set $\tilde c_i^t=0$ if no exchange is attempted.

\begin{table}[!t]
  \centering
  \caption{Per-step reward components used in the return. Positive terms promote progress; penalties encode safety and communication costs.}
  \label{tab:reward-return}
  \renewcommand{\arraystretch}{1.12}\setlength{\tabcolsep}{6pt}
  \begin{tabular}{l l l}
    \hline
    \textbf{Component} & \textbf{Indicator / Term} & \textbf{Weight / Value} \\
    \hline
    Goal completion & $\mathbf{1}\{\text{goal}_i^{t+1}\}$ & $+R_{\text{goal}}$ (e.g., $+100$) \\
    Step penalty & $1$ & $-\,c_{\text{step}}$ (e.g., $0.1$) \\
    \textbf{Transition failure}$^\dagger$ & $\mathbf{1}\{\text{transition fail}_i^t\}$ & $-\,C_{\text{fail}}$ (e.g., $10$) \\
    No-move penalty$^\ddagger$ & $\mathbf{1}\{\text{no-move}_i^t\}$ & $-\,C_{\text{wait}}$ (e.g., $1$) \\
    Distance shaping & $\Delta d_i^t=d_i^t-d_i^{t+1}$ & $+\,\alpha_d\,\Delta d_i^t$ \\
    Comm. fail penalty & $1-\tilde c_i^t$ & $-\,C_{\text{comm}}$ \\
    \hline
  \end{tabular}
  \\[2pt]
  \raggedright\footnotesize
  $^\dagger$ \emph{Transition failure} includes: (i) off-grid or wall/obstacle hit; (ii) \textbf{vertex conflict} (two agents target the same cell); (iii) \textbf{head-on edge conflict} (swap on an edge). \\
  $^\ddagger$ ``No-move'' triggers when the agent remains in place after arbitration (excluded at goal with stay-on-arrival).
\end{table}

As the primary metric, we report the total number of completed tasks (TNCT) over lifelong execution, together with safety and robustness under wireless-link uncertainty.

\subsection{Local Policy}

The on-board controller is a lightweight recurrent local policy
$\pi_{\text{local}} : (o_i^t, h_i^{t-1}) \!\mapsto\! \Delta(\mathcal A)$.
Each agent maintains a GRU hidden state $h_i^{t-1}$ that summarizes its recent
observation history. At every step $t$, the observation $o_i^t$ stacks a
three-channel egocentric FOV (static obstacles, other agents, and a heuristic
guide map), a relative goal vector, and normalized coordinates.
A CNN--GRU encoder produces a hidden state $h_i^t$, from which we derive
policy logits $\ell_i^t$ and a value estimate $V_i^t = v_{\theta_l}(h_i^t)$.

Let $\mathcal A=\{\text{stay},\text{up},\text{down},\text{left},\text{right}\}$
and $m(o_i^t)\in\{0,1\}^{|\mathcal A|}$ be a legality mask that suppresses
out-of-bounds and obstacle-colliding moves.
Upon reaching its goal, we force the agent to remain at its current cell by disabling all move actions in the policy output; concretely, we mask out the logits of every non-stay action so that the policy can only select the stay action.
\begin{equation}
\pi_{\text{local}}(a_i^t \mid o_i^t, h_i^{t-1})
=\frac{\exp(\ell_{i,a}^t)\,m_a(o_i^t)}
       {\sum_{a'\in\mathcal A}\exp(\ell_{i,a'}^t)\,m_{a'}(o_i^t)}
\in \Delta(\mathcal A).
\end{equation}
The local policy $\pi_{\text{local}}$ serves as a default fallback controller
and guarantees safe execution whenever no downlink is available.

\textbf{Training.}
We adopt a two-phase schedule that first builds a reliable on-board fallback
and then adapts it to the lifelong MAPF objective.

\textit{(i) Behaviour-cloning pretraining.}
We warm-start the local policy by minimizing a masked behaviour-cloning loss
on ODrM* expert trajectories. Let $\mathcal D$ be a fixed dataset of triples
$(o_i^t,h_i^{t-1},a_i^{t,*})$ collected under the same legality mask $m(o)$
used at execution time. Here
\[
L_{\mathrm{BC}}(\theta_\ell)
= \mathbb{E}_{(o,h,a^\ast)\sim \mathcal{D}}
\big[-\log \pi_{\theta_\ell}(a_{t,i}^\ast \mid o_t^i, h_{t-1}^i)\big]
\tag{4}
\]
is the standard behavioral cloning loss: the expected negative log-likelihood of the expert action $a_{t,i}^\ast$ under the local policy $\pi_{\theta_\ell}$, given the observation $o_t^i$ and hidden state $h_{t-1}^i$. The policy output is masked so that any illegal action (e.g., stepping into an obstacle or outside the map) is assigned zero probability before the softmax, and thus never appears as a valid label or prediction.

We minimize $L_{\mathrm{BC}}$ on demonstration trajectories generated by ODrM* using mini-batch Adam, obtaining a pretrained parameter vector $\theta_\ell^{(0)}$ that closely imitates ODrM* while always respecting all legality constraints. In addition, optional class reweighting or mild label smoothing can be applied within $L_{\mathrm{BC}}$ to reduce bias towards very frequent actions such as stay. This pretraining stage is crucial for our hybrid design: it yields a strong and safe default controller that remains usable even under severe execution noise and packet loss, so that the cloud component only needs to learn small residual refinements on top of this local baseline.

\textit{(ii) Actor--critic fine-tuning.}
Starting from the behavior-cloned parameters $\theta_\ell^{(0)}$, we further fine-tune the local policy $\pi_{\text{local}}$ with an entropy-regularized advantage actor–critic objective matched to the TNCT reward. The network consists of a shared encoder that maps $(o_t^i, h_{t-1}^i)$ to a latent state $h_t^i$, and two heads on top of this latent state: a policy head that outputs the action distribution $\pi_{\theta_\ell}(\cdot \mid h_t^i)$ and a value head that predicts the state value $V_{\theta_\ell}(h_t^i)$.

Advantages $\hat A_t^i$ are computed with generalized advantage estimation (GAE) using discount factor $\beta \in (0,1)$ and trace parameter $\lambda \in [0,1]$. Given a truncated $\tau$-step return $R_t^i$, the overall fine-tuning loss is
\begin{equation*}
\mathcal L_{\text{AC}}
= \mathbb{E}\!\Big[-\log \pi_{\theta_l}(a_i^t\mid o_i^t,h_i^{t-1})\,\hat A_i^t
\;+\; c_v\big(R_i^t - V_{\theta_l}(h_i^t)\big)^2
\;
\end{equation*}
\begin{equation*}
-\; \alpha\,H\big(\pi_{\theta_l}(\cdot\mid h_i^t)\big)\Big]
\;+\; \lambda_{\mathrm{BC}}\,\mathcal L_{\mathrm{BC}},
\end{equation*}
where $\hat A_t^i$ denotes the GAE, $\alpha \ge 0$ is the entropy weight, $H(\cdot)$ is the Shannon entropy, $c_v \ge 0$ is the value-loss weight, and $\lambda_{\mathrm{BC}} \ge 0$ is an auxiliary behavioral-cloning weight that is gradually annealed towards zero during fine-tuning.

Gradients of $L_{\mathrm{AC}}$ are backpropagated through the shared encoder and both heads, so that the latent representation, the policy, and the value estimates are jointly adapted. For stability, the bootstrap target $R_t^i$ uses a delayed target value network (a slowly updated copy of the encoder and value head) that does not receive direct gradients from $L_{\mathrm{AC}}$. The resulting local policy is explicitly optimized for the long-horizon TNCT objective while remaining stable enough to serve as the backbone of the proposed cloud–edge hybrid controller.

% In this subsection we write $\pi_{\text{local},\theta}\equiv\pi_\theta$, 
% $\ell_{\text{local},i}^t\equiv \ell_i^t$, and $V_{\text{local},\theta}(h_i^t)\equiv V_\theta^t$ for brevity.

\textbf{Uplink allocator.}
At step $t$, a selected sender $i$ uploads a list of vectors including its global location $x_i^t\!\in\!\mathbb R^2$, locations of peers in its FOV $\{x_j^t\}_{j\in \mathrm{FOV}_i^t}$,
and recurrent hidden states $\{h_j^t\}_{j\in \{i\}\cup \mathrm{FOV}_i^t}$ of
dimension $d_h$ 
%%%%%%%%%%no assumptions

Let $U$ denote the set of agents. In addition, we pre-identify a set $C$ of \emph{risk centers} in the environment, such as grid cells near narrow corridors, doors, and busy intersections where vertex and edge conflicts or long queues are likely to occur. For each risk center $c \in C$, we assign a nonnegative event-risk weight $w_c \ge 0$ that summarizes the impact or likelihood of congestion at that location. For any $c \in C$, we define $\mathrm{NN}(c) \subseteq U$ as the subset of agents whose current field of view (FOV) covers $c$, i.e., agents that can observe and report local information about this risk center. At each time step $t$, we then select a sender set $S_t \subseteq U$ with $|S_t| \le K$ to maximize

\[
F_t(S)=\sum_{u} w_u\!\left(1-\prod_{i\in S\cap \mathrm{NN}(u)}\big(1-\tilde p^{\mathrm{UL}}_{i}\big)\right),
\]
where $\tilde p^{\mathrm{UL}}_{i}$ is a nominal uplink success probability obtained from the finite-blocklength model under a baseline resource-block assignment. For a given risk center $u$, the inner term
\[
1-\prod_{i\in S\cap \mathrm{NN}(u)}(1-\tilde p^{\mathrm{UL}}_{i})
\]
is the probability that at least one selected sender successfully uploads in this step (under independent UL success events), instead of merely checking whether $u$ is covered by some sender. Thus $F_t(S)$ extends a plain weighted-coverage score that would only count whether $u$ is covered at all, by also accounting for the physical-layer reliability of each uplink via $\tilde p^{\mathrm{UL}}_{i}$.

Under the standard independence assumption on UL success events, this set function $F_t$ is monotone submodular, and the classical greedy selection rule for a cardinality-constrained problem $\lvert S\rvert \le K$ achieves a $(1-1/e)$ approximation guarantee to the optimal value~\cite{nemhauser1978analysis,krause2014submodular}.

% \paragraph{Bitrate-aware RB assignment.}
Given the sender set $S_t$, we first compute each selected agent’s uplink message bit-length $b_i^{\mathrm{UL},t}$ and sort $i \in S_t$ by $b_i^{\mathrm{UL},t}$. For each agent $i$, the scheduler then assigns $m_i^t$ uplink resource blocks (RBs); each RB provides $n$ channel uses, so the total blocklength allocated to $i$ at step $t$ is
\[
n_i^t = m_i^t\, n.
\]
The corresponding code rate is
\[
R_i^t = \frac{b_i^{\mathrm{UL},t}}{n_i^t},
\]
which we aim to keep within the target MCS. Under a single-band constraint, each scheduled agent occupies at most one band, but can still receive more or cleaner RBs (larger $m_i^t$ or higher-SNR RBs) on that band.

In the finite-blocklength regime, for fixed SNR $\gamma$ and blocklength $n_i^t$, the block error probability is an increasing function of the rate $R_i^t$. Therefore, agents with longer messages are preferentially granted more or higher-SNR spectrum whenever possible. This allocation is limited by the uplink RB budget
\[
\sum_i m_i^t \le C_{\mathrm{UL}},
\]
where $C_{\mathrm{UL}}$ is the total number of RBs available on the uplink in one control cycle. When this budget is tight (i.e., $\sum_i m_i^t = C_{\mathrm{UL}}$), no additional RBs can be assigned, so some $R_i^t$ must increase and their corresponding packet error rates $\mathrm{PER}_i^t$ also increase.

\subsection{Cloud Policy and Event-Aware Downlink Allocation}

\textbf{Cloud prior and residual.}
From the uplink, the cloud receives for each agent $i$ a compact prior
$z_t^i = (\{h_t^j\}, \{x_t^j\})$ that contains, for all agents $j$ within $i$’s FOV, their GRU hidden states $h_t^j$ and current grid positions $x_t^j$. Each hidden state $h_t^j$ is the output of the on-board GRU encoder (parameterized by $\theta_\ell$) and summarizes agent $j$’s recent observation and motion history in a fixed-size vector.

On the server we instantiate a mirrored copy of the on-board local policy network, consisting of the same GRU encoder and policy head with parameters tied to $\theta_\ell$. From $z_t^i$ we extract agent $i$’s own hidden state $h_t^i$ and feed it through this mirrored policy head to reconstruct the local action logits
\[
\ell_{t,i}^{\mathrm{loc}} \in \mathbb{R}^{|\mathcal{A}|},
\]
which are exactly the logits that the on-board policy would produce for agent $i$ in the absence of cloud refinement.

On top of this frozen mirrored network, we attach a small residual head $a_\psi(\cdot)$ with parameters $\psi$. This head takes as input the prior $z_t^i$ (and optionally $\ell_{t,i}^{\mathrm{loc}}$) and outputs a correction
\[
\Delta \ell_t^i = a_\psi(z_t^i, \ell_{t,i}^{\mathrm{loc}}) \in \mathbb{R}^{|\mathcal{A}|}.
\]
The cloud-refined logits are then defined as
\[
\ell_{t,i}^{\mathrm{cloud}} = \ell_{t,i}^{\mathrm{loc}} + \Delta \ell_t^i,
\]
so the cloud only learns to adjust the local policy by a residual term rather than replacing it entirely.

The cloud-side action proposal that may be sent on the downlink is therefore
\[
\hat\ell_t^i = \ell_{t,i}^{\mathrm{cloud}},
\]
which is later gated and, if selected, transmitted in the downlink 
(see Fig.~\ref{fig:BD}).

\textbf{Fused context and encoders.}
The cloud fuses UL priors into a global context that is cropped around each
agent.
(i) Global positions $\{x_j^t\}$ are rasterized into a heuristic map $M_{heu,c}^t$ and
per-agent PER maps $M_{PER,c}^t$ using the finite-blocklength channel model.
(ii) The mirrored GRU rolls out $N$ steps under the local policy to obtain
predicted logits, which are converted by a stochastic kernel into a
short-horizon belief map $\mathbf B_{i,t}$.
(iii) Each agent extracts its egocentric crop and feeds three lightweight CNN
encoders: Enc-1 (heuristic + PER + belief + estimated logits) outputs
$h_i^{\text{pre}}$ for the actor residual; Enc-2 (belief + GRU logits) outputs
$h_i^{\text{gru}}$ for the relief head; Enc-3 (EEM features) outputs
$h_i^{\text{eve}}$ for the risk head.

\textbf{Event estimator (EEM).}
Event estimator (EEM). For each agent $i$ and step $t$, the EEM takes as input the same egocentric fused crop used in the previous section: a multi-channel patch centered at $x_t^i$ that stacks the heuristic map, the UL PER map, the short-horizon belief map derived from the mirrored GRU logits, and an optional occupancy channel. From this crop it predicts, for each event type $e \in \mathcal{E} = \{\text{vertex}, \text{edge}, \text{wall}, \text{wait}\}$, a probability
\[
\hat y_{t,i,e} \in [0,1]
\]
that the event will occur in the near future.

To obtain training targets $y_{t,i,e}$, we perform Monte Carlo rollouts of length $H$ from the current state, following the current local policy and simulating all stochastic effects (transition noise, conflicts, communication outcomes). For each rollout we record whether event $e$ occurs at step $\tau \in \{1,\dots,H\}$ and assign it a time-decay factor $\xi^\tau$ with $\xi \in (0,1]$. Averaging these discounted indicators over rollouts yields an estimate of the short-horizon probability that event $e$ will occur if we keep following the current local policy. The EEM is then trained with a class-balanced binary cross-entropy loss,
\[
L_{\mathrm{EEM}} = \mathbb{E}\Bigg[\sum_{e\in\mathcal{E}} \alpha_e \, \mathrm{BCE}\big(\hat y_{t,i,e},\, y_{t,i,e}\big)\Bigg],
\tag{6}
\]
where $\alpha_e$ reweights event types to compensate for class imbalance.

For the \emph{relief} supervision, we additionally construct counterfactual pairs of rollouts from the same starting state. In the “no-cloud” rollout (branch $(0)$), agents act purely according to the local policy. In the “with-cloud” rollout (branch $(1)$), the hybrid controller is used and cloud residual actions are applied whenever downlink is available. This yields two sets of event targets $y^{(0)}_{t,i,e}$ and $y^{(1)}_{t,i,e}$, and we define the counterfactual risk reduction
\[
\Delta y_{t,i,e} = y^{(0)}_{t,i,e} - y^{(1)}_{t,i,e},
\]
which quantifies how much the probability of event $e$ decreases when cloud assistance is enabled. The relief head is trained to predict these deltas from the fused crop. For stability, gradients from this relief loss are not backpropagated through the actor parameters used in the “with-cloud” branch (1); the policy in that branch is treated as fixed when constructing the targets.

\textbf{Heads and selector.}
The risk head $W$ takes the event embedding $h_i^{\mathrm{eve}}$ (from Enc-3) and maps it to per-event risk urgencies $\rho_{t,i,e} \in [0,1]$ for each $e \in \mathcal{E}$. A large value of $\rho_{t,i,e}$ indicates that, given the local EEM features, event $e$ (vertex conflict, edge conflict, wall hit, or wait) is both likely and important around agent $i$ in the near future. The relief head $P$ takes the GRU-based embedding $h_i^{\mathrm{gru}}$ (from Enc-2) and outputs per-event relief scores $\sigma_{t,i,e} \in [0,1]$, which are trained to match the counterfactual risk reductions $\Delta y_{t,i,e}$. Thus $\sigma_{t,i,e}$ measures how much the risk of event $e$ can be reduced if cloud assistance is applied around agent $i$. The actor head $a_\psi$ maps the fused context embedding $h_i^{\mathrm{pre}}$ (from Enc-1) to residual action logits $\Delta \ell_t^i = a_\psi(h_i^{\mathrm{pre}})$ that refine the local policy.

We combine risk and relief to form a scalar scheduling score for each agent:
\[
s_t^i = \sum_{e\in\mathcal{E}} \omega_e \, \rho_{t,i,e}\, \sigma_{t,i,e},
\tag{7}
\]
where the nonnegative weights $\omega_e$ specify the relative importance of different event types (e.g., vertex and edge conflicts can be given higher weights than wall hits or waits). Intuitively, $s_t^i$ is large when agent $i$ is in a high-risk region ($\rho_{t,i,e}$ large) and the cloud is expected to provide substantial relief there ($\sigma_{t,i,e}$ large).

At each step we sort all agents by $s_t^i$ and mark the top $C_{\mathrm{DL}}$ agents as eligible for downlink. For each agent $i$ we then define a gating variable $g_t^i$ that determines how strongly the cloud residual is applied. Here we use a binary indicator $g_t^i = \mathbf{1}\{i \text{ is selected in the top } C_{\mathrm{DL}}\}$, so only selected agents receive cloud-refined actions while others rely solely on the local policy.

\textbf{Conflict gating.}
Every $H$ steps, we randomly assign each agent a global priority $\kappa_i^t$ by sampling according to its scheduling weight (e.g., proportional to $s_t^i$), which induces a total ordering over agents that is then restricted to each conflict cluster to provide a consistent priority ordering.

In the fused belief for agent $i$, predicted footprints of higher-priority
agents ($\kappa_j^t<\kappa_i^t$) are treated as hard obstacles, while
footprints of lower-priority agents are ignored, creating asymmetric guidance
and right-of-way corridors.
We also down-weight high-priority agents in the scheduler by multiplying
$s_i^t$ with a decreasing function $\phi(\kappa_i^t)$, so lower-priority agents
are more likely to receive downlink refinements.
Together, priority-aware masking and scheduling implement a soft centralized
``who moves first'' rule that encourages cooperative collision avoidance.

\textbf{Cloud training.}
During cloud training we freeze the local parameters and optimize the residual
policy, allocator, EEM and auxiliary heads on on-policy trajectories.
We reuse the same per-step reward $r_i^t$ and the same discount $\beta$,
GAE parameter $\lambda$ and advantage horizon $\tau$ as in the local training;
advantages $\hat A_i^t$ are computed with GAE.
To discourage overly aggressive residuals we define a stability regularizer
based on the Kullback–Leibler divergence between the local and executed action
distributions,
\begin{equation}
\Delta\mathcal D_i^{t,\mathrm{eff}}
= D_{\mathrm{KL}}\!\Big(\mathrm{Softmax}(\ell_i^t)\,\big\|\,
\mathrm{Softmax}(\ell_i^t+g_i^t\Delta\ell_i^t)\Big).
\end{equation}
The cloud loss combines a policy term, a value term, the stability regularizer
and the two EEM-supervised head losses:
\begin{equation*}
% \begin{align*}
\mathcal L_{\mathrm{cloud}}
= \mathbb E\Big[-\,w_i^t\log \pi_{\mathrm{cloud}}(a_i^t\mid z_i^t)\,\hat A_i^t
-\alpha_{\mathrm c}\,H\big(\pi_{\mathrm{cloud}}(\cdot\mid z_i^t)\big)
\end{equation*}
\begin{equation*}
\quad +\,c_v^{\mathrm c}\big(R_i^t - V_{\psi}(z_i^t)\big)^2\Big]
+ \lambda_{\mathrm{KL}}\,\mathbb E\big[\Delta\mathcal D_i^{t,\mathrm{eff}}\big]
+ \lambda_W\,\mathcal L_W + \lambda_P\,\mathcal L_P,
% \end{align*}
\end{equation*}
where $w_i^t\in[0,1]$ encodes downlink success or predicted reliability,
$\alpha_{\mathrm c}\!\ge\!0$ is the entropy weight,
$H(\cdot)$ denotes Shannon entropy, $c_v^{\mathrm c}\!\ge\!0$ is the
value-loss weight, and $\lambda_{\mathrm{KL}},\lambda_W,\lambda_P\!\ge\!0$ are
tunable coefficients.
Here $\mathcal L_W$ and $\mathcal L_P$ are simple squared-error losses that
fit $\rho_{i,e}^t$ to $\hat y_{i,e}^t$ and $\sigma_{i,e}^t$ to
$\Delta\hat y_{i,e}^t$, respectively.
Gradients update the residual policy parameters $\psi$, the allocator (which
sets $g_i^t$), the EEM and the auxiliary heads, while the local GRU and policy
head remain frozen.

\textbf{Downlink allocation.}
At runtime the downlink allocator uses the same ingredients to score each
agent:
\begin{equation}
s_i^{t,\mathrm{DL}}
=\phi(\kappa_i^t)\; p_i^{\mathrm{DL},t}\;
\Big(\textstyle\sum_{e\in\mathcal E}\omega_e\,\rho_{i,e}^t\,\sigma_{i,e}^t\Big)\;
\Delta\mathcal D_i^t,
\end{equation}
where $p_i^{\mathrm{DL},t}$ is the predicted downlink success probability and
$\Delta\mathcal D_i^t = D_{\mathrm{KL}}(\mathrm{Softmax}(\ell_i^t)\,\|\,
\mathrm{Softmax}(\ell_i^t+\Delta\ell_i^t))$ measures how much the cloud
residual would change the local action distribution.
The top-$C_{\mathrm{DL}}$ agents according to $s_i^{t,\mathrm{DL}}$ are granted
downlink.

When the downlink message length scales with the number of peers in the
sender's FOV we write
$b_i^{\mathrm{DL},t}=b_0+b_1|\mathrm{FOV}_i^t|$.
With a fixed per-packet blocklength $n$, the code rate
\begin{equation}
R_i^{\mathrm{DL},t}=\frac{b_i^{\mathrm{DL},t}}{n}
\end{equation}
increases with $|\mathrm{FOV}_i^t|$, which in the finite-blocklength regime
raises the block error probability.
The allocator therefore prefers to assign higher-SNR RBs to heavier, high-score
messages, effectively trading rate for reliability.

\vspace{0.25em}
% \noindent\text{Remark.} The formulation is agnostic to the exact event taxonomy and risk weights; any calibrated EEM that outputs per-action risk can be plugged into~\eqref{eq:action-risk}--\eqref{eq:relief-score} without changing the guarantees or training recipe.

% \subsection{Training Strategy}

\section{Numerical Simulation}

\subsection{Environment and Metrics}

We evaluate on standard MovingAI scenarios \cite{sturtevant2012benchmarks} covering four representative
topologies: Empty ($48{\times}48$), Random ($32{\times}32$), Maze
($32{\times}32$), and den312d ($65{\times}81$).
Unless stated otherwise, each episode lasts $H{=}128$ steps and each agent
uses an egocentric $7{\times}7$ FOV with three channels (obstacles, other
agents, self).
Our primary metric is the \emph{Total Number of Completed Tasks} (TNCT),
i.e., the cumulative number of start--goal tasks finished by all agents within
the horizon.
With fixed $H$ and throughput defined as tasks per step,
\begin{equation}
\text{TNCT} \;=\; H \times \text{Throughput}.
\end{equation}
TNCT directly answers the operational question ``how many jobs finish by the
deadline,'' aligns with real KPIs (e.g., orders dispatched), and is insensitive to implementation details such as the choice of control step frequency or the particular reward-shaping weights used during training. We use preset MovingAI scenarios and fixed random seeds to sample start–goal pairs.

Action execution follows the stochastic kernel $K_{\varepsilon}$ defined in
Sec.~\ref{MAPF Problem Statement}, which redistributes a small probability mass from the intended forward
cell to stay/side/back outcomes.
Unless stated otherwise we instantiate
\(\varepsilon = (\varepsilon_{\mathrm{stay}},\varepsilon_{\mathrm{side}},
\varepsilon_{\mathrm{back}}) = (0.05, 0.05, 0)\),
so that an intended cardinal move succeeds with probability $0.90$, stays
with probability $0.05$, and the remaining $0.05$ is split equally between
the two side cells.
For the stay action $(0,0)$ the kernel collapses to a deterministic stay,
and if a sampled destination is a wall or outside the grid we apply the same
bounce-to-stay rule as in Sec.~\ref{MAPF Problem Statement}.

Before state updates we record attempted interactions---wall
(blocked/out-of-bounds), vertex (same target cell), edge-swap (position
exchange), and moves into a staying agent.
A safety arbitration layer then enforces non-penetration: wall and edge-swap
attempts are forced to stay; vertex conflicts admit a single winner; moves
into a staying agent are blocked.
We expose fine-grained wait causes (goal, intent, wall, vertex conflict, edge
conflict) and a boolean transition-failure flag for training and diagnostics.
This unified execution model serves three purposes. First, it captures near-field interactions where conflicts arise between neighboring agents. Second, it injects controlled execution noise that penalizes brittle policies while still keeping the tasks solvable. Third, it mirrors practical controllers, where safety rules arbitrate the final executed actions and the learned policies focus on reducing risky intents.

\subsection{Baselines and Hybrid Design}

We consider $N$ agents and $K_o<N$ orthogonal downlink channels, with
channel ratio $r_{\mathrm{ch}} \triangleq K_o/N$.
Let $\gamma_i(t)$ denote agent $i$'s instantaneous SNR.
At each step we select the top-$K_o$ agents by SNR,
\begin{equation}
\mathcal{K}(t)=\operatorname*{Top\text{-}K_o}_{i\in\{1,\dots,N\}}\!\big(\gamma_i(t)\big), \qquad
c_i(t)=\mathbf{1}\{i\in \mathcal{K}(t)\},
\end{equation}
and model reliable packet reception via a finite-blocklength success map
\begin{equation}
\tilde c_i(t)\sim \mathrm{Bernoulli}\!\big(c_i(t)\,q(\gamma_i(t))\big), \qquad q(\gamma)\in(0,1].
\end{equation}

\emph{Communication-aware centralized baseline.}
We also construct a centralized baseline that still respects the communication model. When agent $i$ has no successful UL--DL exchange at step $t$ ($\tilde c_t^i = 0$), it simply follows its precomputed local A* guidance and executes the next move along its path,
\[
a_i(t) = A^*(s_i(t)).
\]
When a UL--DL exchange succeeds ($\tilde c_t^i = 1$), control is handed over to a centralized planner. We run a localized instance of ODrM on the subset $V^t$ of connected agents that lie within a $9\times 9$ window around any detected conflict, using their current states $s_{V^t}(t)$ as the initial condition. ODrM is run in anytime mode with a strict wall-clock time budget $\tau$ per control step, and the resulting centralized action for agent $i$ replaces the A* move:
\[
a_i(t) =
\begin{cases}
A^*(s_i(t)), & \tilde c_t^i = 0,\\[3pt]
\text{ODrM}_{\text{local},\tau}(s_{V^t}(t))_i, & \tilde c_t^i = 1.
\end{cases}
\tag{14}
\]
The time budget $\tau$ models real-time timing constraints: if the conflict region $V^t$ is large, ODrM may not fully converge within $\tau$, so the centralized plan is truncated and can be suboptimal.

This baseline remains communication-aware. When the channel is rich ($r_{\mathrm{ch}} = 1$, so all agents are mutually visible) and the finite-blocklength success probability $q(\gamma)$ is close to one, most agents are connected at each step and the localized ODrM instance effectively covers the whole team, so the behavior approaches running a global ODrM planner at every time step. At realistic SNRs and with finite $\tau$, however, the baseline still experiences occasional packet loss and incomplete centralized replanning.

Because communication is modeled as a first-class constraint through $r_{\mathrm{ch}}$ and the finite-blocklength success $q(\gamma)$, we do not compare against an oracle fully centralized planner that assumes perfect, cost-free communication at every step. Instead, we use two interpretable reference points: the communication-aware ODrM+A* scheme above and a purely local A* controller. The former acts as an upper reference that exploits centralized coordination under the given communication limits, while the latter is a robust lower reference that uses no communication at all. Our hybrid cloud–edge policy is designed to track the centralized reference when links are reliable, while never degrading below the local A* performance when bandwidth is scarce.

\begin{figure*}[t]
  \centering
  \subfloat[Random (32$\times$32)\label{fig:bands-32}]{
    \includegraphics[width=0.227\textwidth]{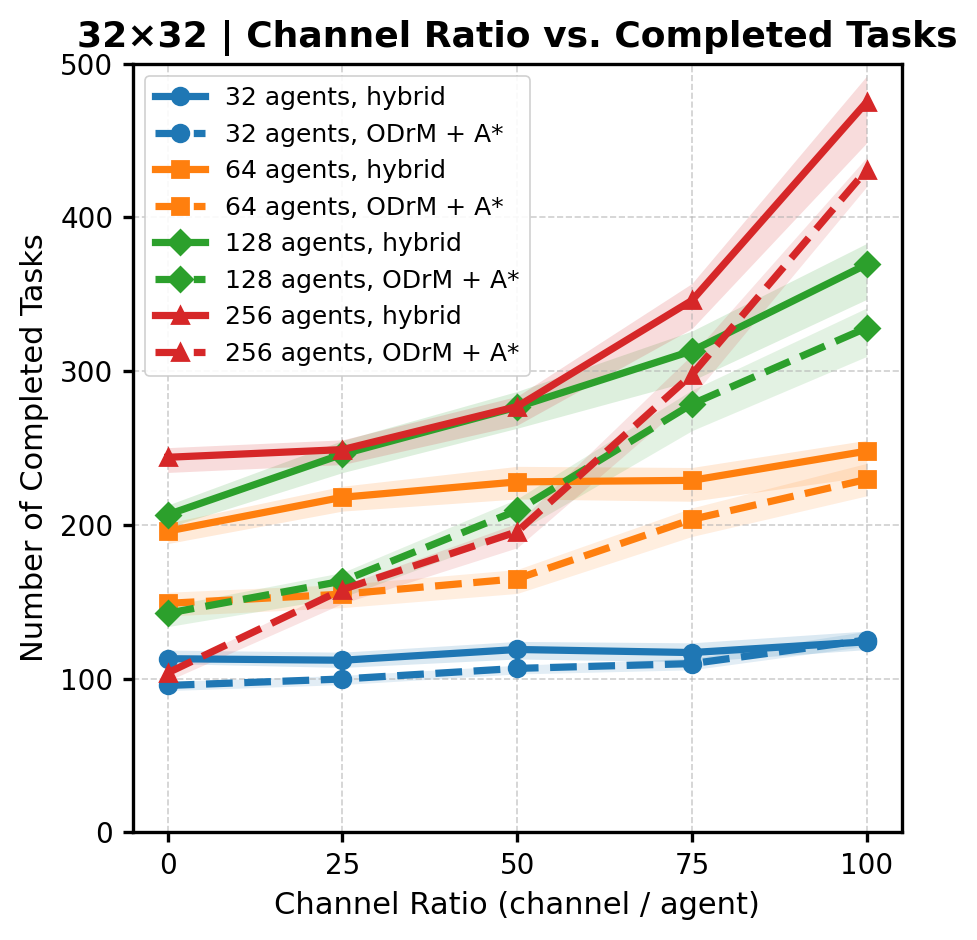}
  }\hfill
  \subfloat[Random (64$\times$64)\label{fig:bands-64}]{
    \includegraphics[width=0.227\textwidth]{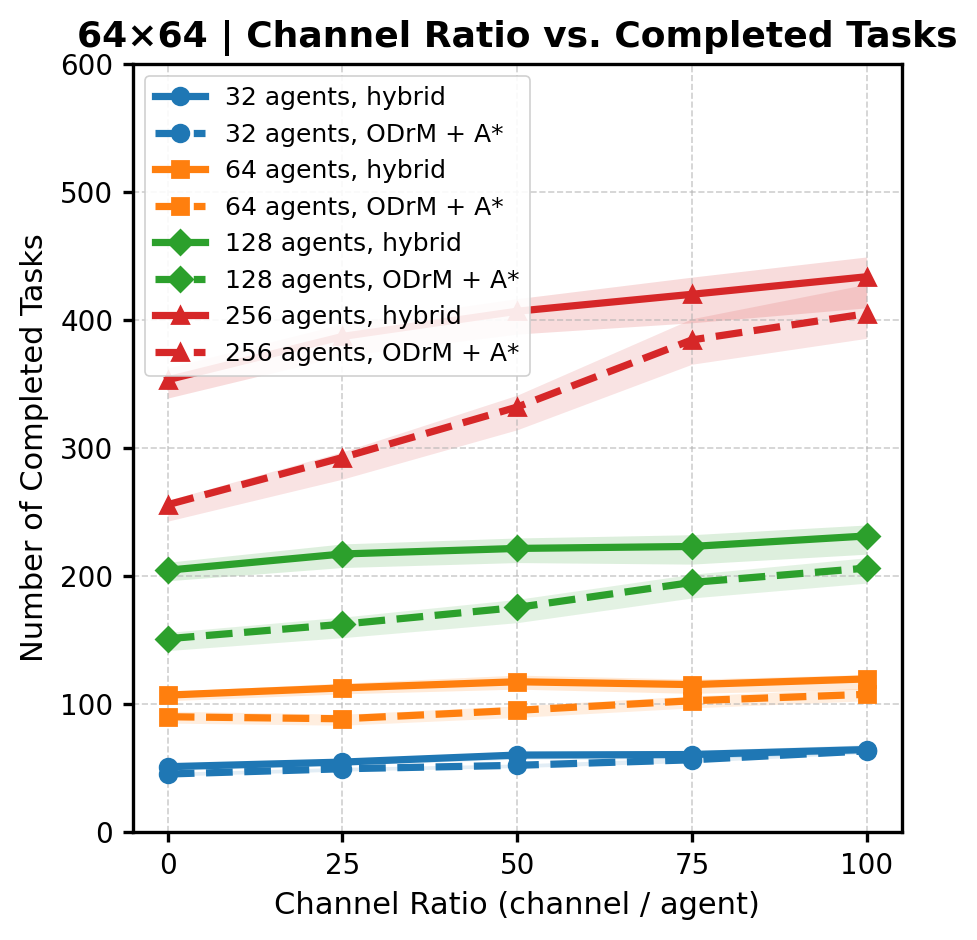}
  }\hfill
  \subfloat[Maze (32$\times$32)\label{fig:bands-maze}]{
    \includegraphics[width=0.24\textwidth]{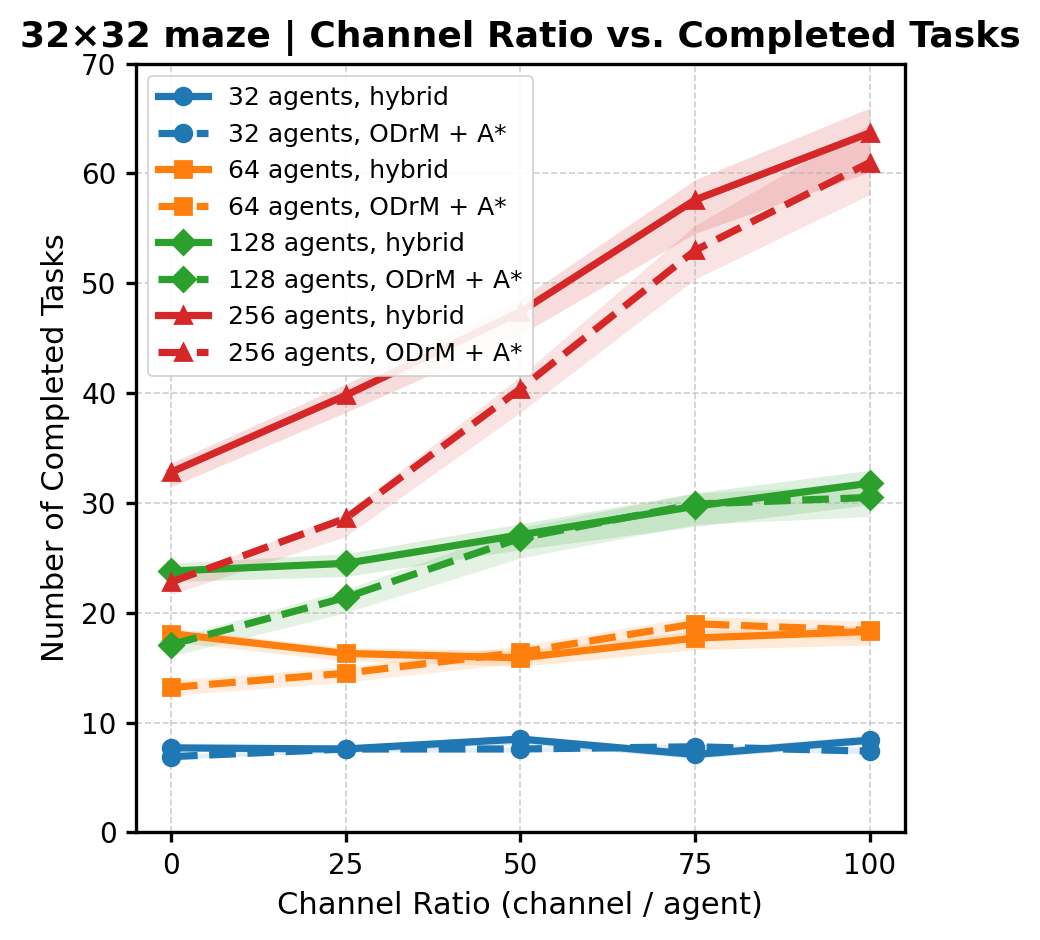}
  }\hfill
  \subfloat[den312d (65$\times$81)\label{fig:bands-312d}]{
    \includegraphics[width=0.227\textwidth]{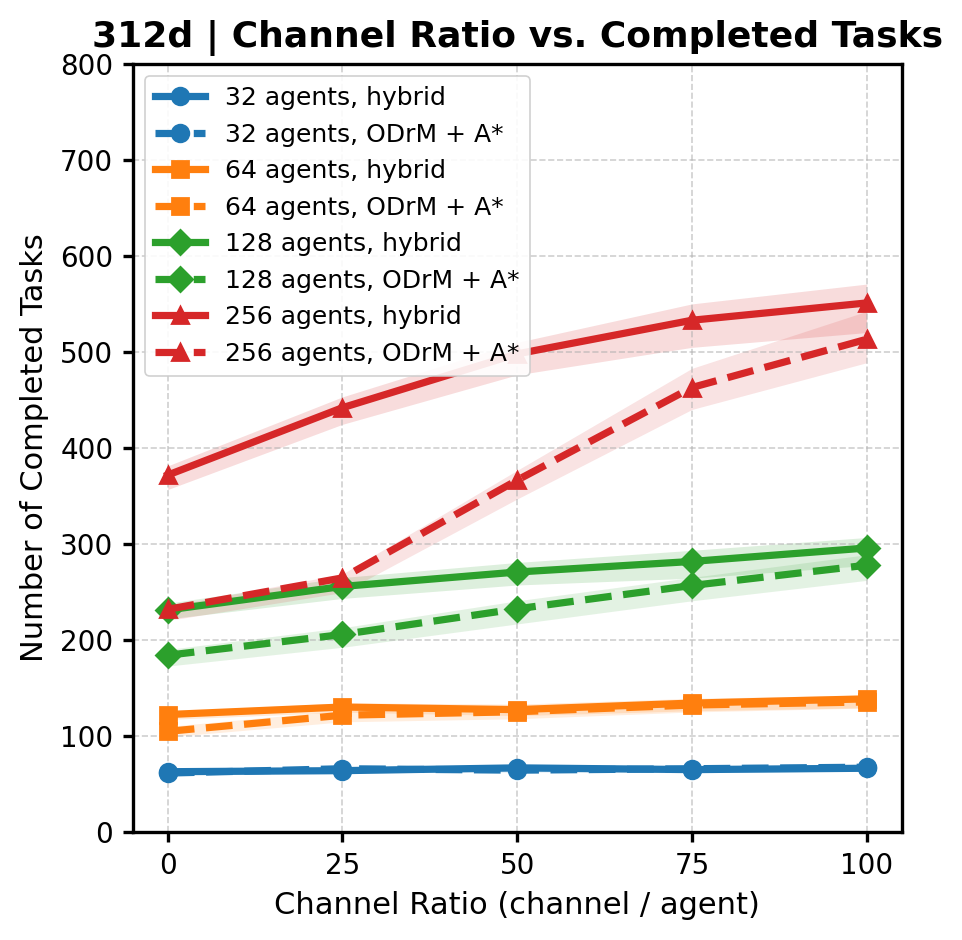}
  }
  \caption{TNCT (128-step horizon) across MovingAI maps with
  $N\!\in\!\{32,64,128,256\}$ and channel ratio $r_{ch} \!\in\!\{0,25,50,75,100\}$,
  under a unified execution model: egocentric $7{\times}7$ FOV, a 0.90-centered
  stochastic transition kernel, cloud overriding local upon packet reception,
  and safety via attempt-arbitration.}
  \label{fig:ood-MovingAI}
\end{figure*}

\subsection{Throughput Under Bandwidth and Congestion}

Fig.~\ref{fig:ood-MovingAI} reports TNCT over a 128-step horizon across
MovingAI maps with $N\!\in\!\{32,64,128,256\}$ and channel ratios
$r_{ch}\!\in\!\{0,25,50,75,100\}\%$.
TNCT increases with $r_{ch}$ on all maps, but exhibits clear diminishing
returns at high bandwidth.
The proposed hybrid controller consistently matches or surpasses
ODrM{+}A* under the same $r_{ch}$, and the gap widens under heavier loads and
stronger structural bottlenecks, as visible on the Maze and den312d maps
in Fig.~\ref{fig:bands-maze} and \ref{fig:bands-312d}.
These curve shapes align with congestion $\rho=N/|{\cal F}|$ and the
pre-arbitration conflict rate
$C=\mathbb{E}[\mathrm{vertex}+\mathrm{swap}]$:
when $\rho$ and $C$ are low (e.g., Random $64{\times}64$ in
Fig.~\ref{fig:bands-64}), the system is communication-limited and TNCT grows
nearly linearly with $r_{ch}$; when obstacles and chokepoints elevate $C$,
the hybrid policy achieves higher TNCT-per-packet by steering scarce cloud
refinements to cluttered pockets and sequencing flows at bottlenecks, often
showing a knee around $r_{ch}\!\in\![50,75]\%$.

Because non-penetration is enforced deterministically by the
attempt--arbitration mechanism, performance differences arise from the
\emph{intent distributions} before arbitration.
Hybrid’s cloud refinement---guided by kernel-induced belief and EEM
risk---shifts probability mass away from wall/vertex/swap intents, reducing
cancellations; as $\rho$ or structural bottlenecks increase $C$, this
risk-aware shaping turns the same number of packets into larger TNCT gains.
When $C\!\approx\!0$ (notably parts of Fig.~\ref{fig:bands-64}), the local
fallback already completes most short jobs and returns diminish with
$r_{ch}$, matching the design goal of ``local first, cloud when it helps.''

\subsection{Runtime Scaling and Ablations}

%#######
Figure ~\ref{fig:ood-MovingAI} focuses on TNCT on a single structured MovingAI map with a moderate team size, which allows us to inspect qualitative behaviour and ablation effects under controlled conditions. However, real deployments also care about how fast a controller can react and how its runtime scales with map size and bandwidth.

\begin{figure}[t]
  \centering
  \subfloat[Performance comparison]{
    % \vspace{10mm}  
    \includegraphics[height=0.2\textwidth]{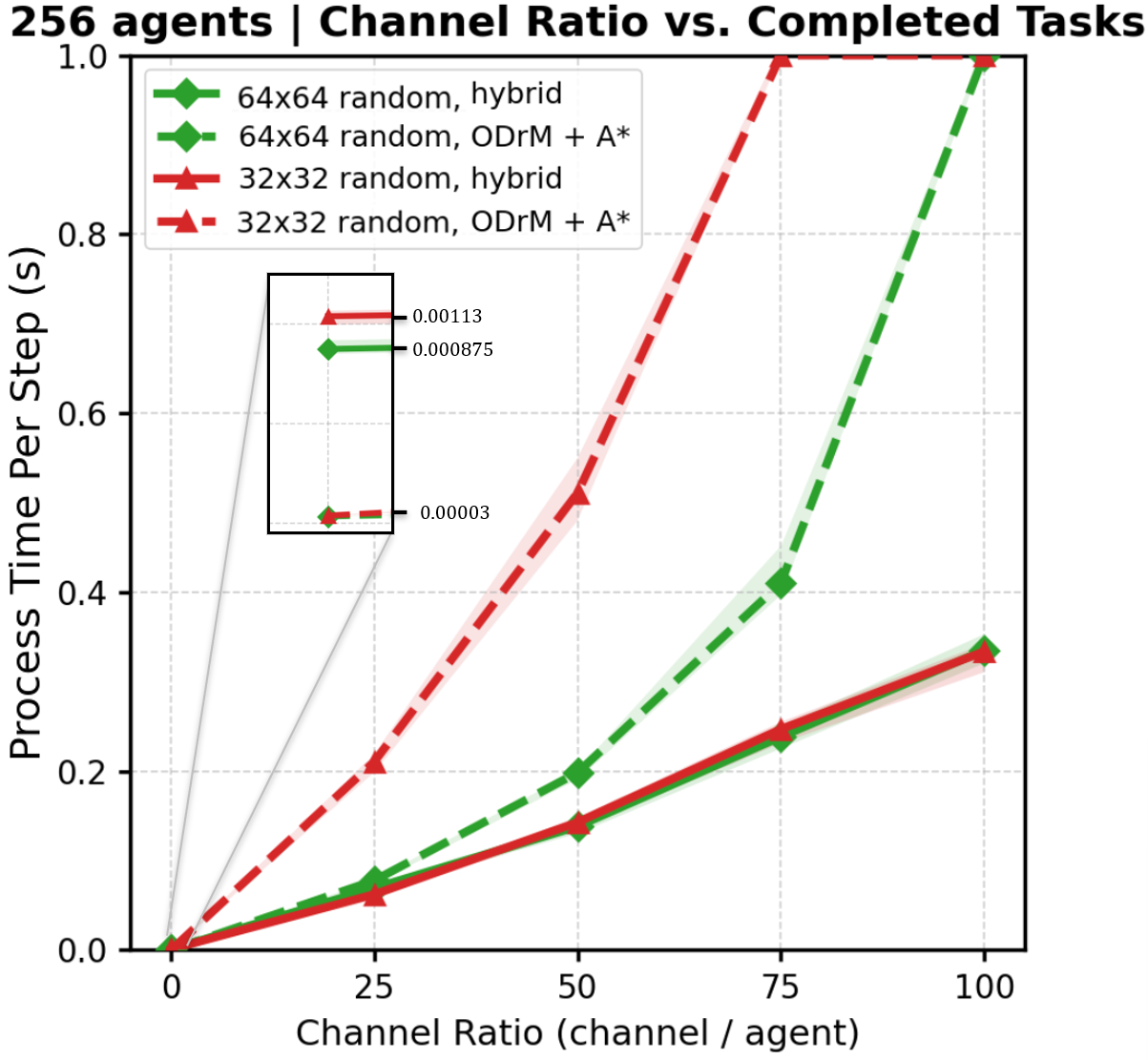}
    \label{fig:abl-channel}
  }\hfill
  \subfloat[TNCT during training]{
    \includegraphics[height=0.21\textwidth]{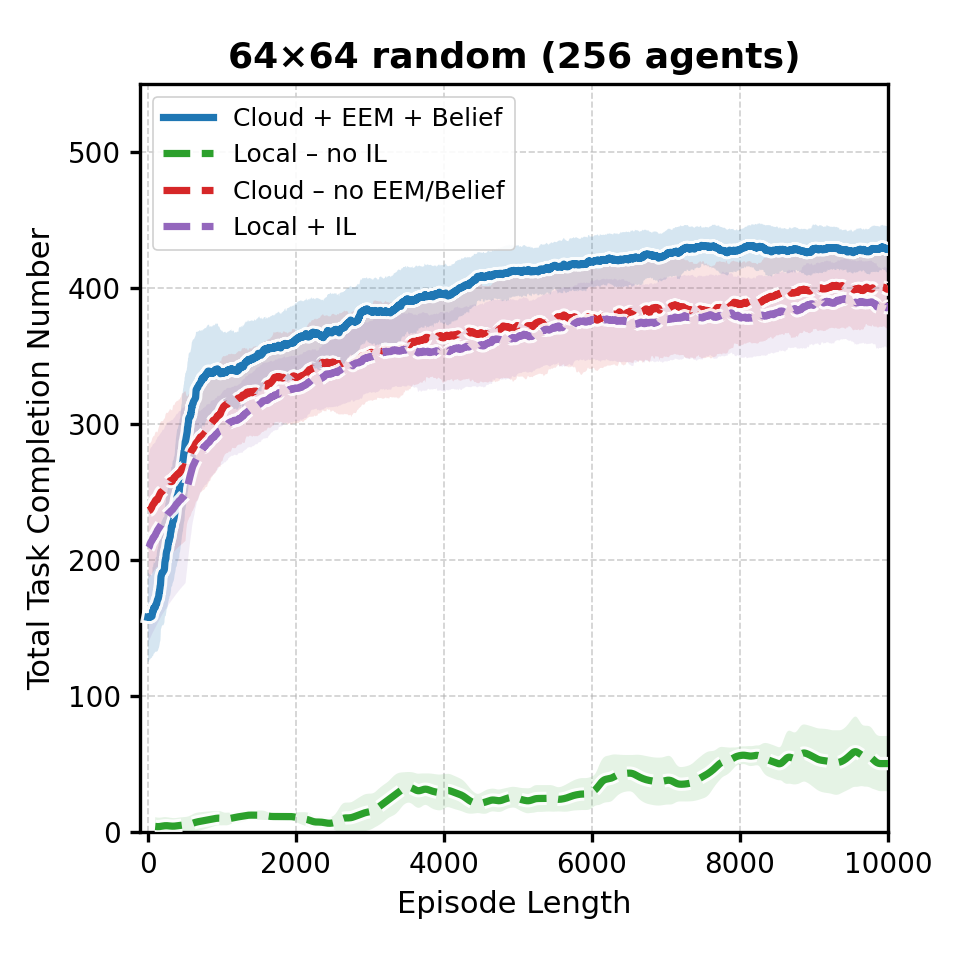}
    \label{fig:abl-train}
  }
  \caption{Performance and training of the hybrid controller.
(a) TNCT versus channel ratio on random MovingAI maps with 256 agents;
the hybrid policy matches or exceeds the communication-aware ODrM{+}A*
baseline at all bandwidths.
(b) TNCT during cloud training on a $64{\times}64$ random map; adding
EEM{+}belief yields the largest gain over local-only and ablated variants.}
  \label{fig:compare}
\end{figure}

To assess runtime scalability, Fig. ~\ref{fig:compare}(a) reports the average wall-clock process time per control step as a function of the channel ratio $r_{\mathrm{ch}}$ on random MovingAI maps with $N=256$ agents and two map sizes (32×32 and 64×64). Across all bandwidths, the hybrid controller remains substantially faster than the communication-aware ODrM+A* baseline; the gap becomes especially pronounced on the larger 64×64 map as both the map and the number of centralized replans grow.

This behaviour reflects both algorithmic and implementation effects. On a 4-connected lattice with $|V| = L^2$ cells, a single A* call incurs $O(|V|\log|V|)$ work. In our implementation of ODrM+A*, repeated replans under agent density $\rho$ lead to an empirical per-step cost that grows roughly like
\[
T_{\text{ODrM}} \approx k(N,\rho)\,|V|\log|V|,
\]
with $k(N,\rho)\ge N$ as conflicts become frequent. This expression is intended as a scaling heuristic rather than a tight asymptotic bound, and is consistent with the observation in~[7] that the computational cost grows with the size of the largest set of colliding robots rather than with the total number of agents. Deriving a formal bound as an explicit function of $N$, $\rho$ and $|V|$ would require strong assumptions on the distribution of conflict clusters and replan triggers, which in MAPF are policy- and map-dependent; worst-case analysis is dominated by NP-hard instances and would not be predictive for the real-time regime we evaluate.

By contrast, the hybrid controller’s per-step cost can be written as
\[
T_{\text{Hybrid}} \approx N C_{\text{loc}} + r_{\mathrm{ch}} N C_{\text{cloud}} + T_{\text{comm}},
\]
where the local pass $C_{\text{loc}}$ operates on an egocentric 7×7 window (constant in $L$) and only $M = r_{\mathrm{ch}} N$ agents receive cloud refinement. This yields near-linear scaling in $N$ and only weak dependence on map size, consistent with the close spacing of the hybrid curves between 32×32 and 64×64 in Fig. 4(a). At $r_{\mathrm{ch}}=0$ the system reduces to a fully decentralised local policy; at $r_{\mathrm{ch}}=100\%$ it behaves like a centralized refinement for all agents, and tuning $r_{\mathrm{ch}}$ offers a natural knob to trade refinement breadth for latency under real-time control budgets.

Fig. ~\ref{fig:compare}(b) reports ablations on random 64×64 maps with $N=256$.
We compare four variants: full Cloud+EEM+Belief, Cloud without
EEM/Belief, Local with imitation learning (IL), and Local without
IL. Across channel ratios, IL lifts the floor of reliable edge
autonomy: Local+IL significantly outperforms Local without IL,
which struggles to escape sparse-reward plateaus under the strict
attempt-then-arbitrate execution model, where many proposed moves
are cancelled by conflict resolution.

EEM and belief account for most of the cloud-side gain: removing
them largely collapses the benefit of the residual and the curve
approaches Local+IL, whereas the full model maintains higher TNCT
and smaller variability in TNCT across random seeds and map
instances (narrower shaded bands in Fig.~4(b)). Together, these
trends suggest a clear division of labor: IL provides a competent
edge fallback, while EEM+belief convert sparse task signals into
dense, interpretable risk indicators over vertex, edge-swap, wall
and wait events, allowing the cloud residual to shift probability
mass away from risky intents earlier and more reliably.

%%%%%%%%%%%%%%%%%%%

\section{Conclusion}

We studied lifelong multi-agent path finding with transition and
communication uncertainties using a cloud--edge hybrid controller.
A lightweight on-board GRU policy provides default autonomy, while a
cloud residual refines actions only when packets arrive, under a unified
execution model (egocentric FOV, stochastic kernel, attempt--arbitration
safety) and TNCT as the primary metric.
An event estimator (EEM), kernel-induced belief, and conflict-aware
downlink allocator convert sparse completions into event-level risk
signals and bandwidth-aware residual corrections.

Across four MovingAI maps and up to 256 agents, the hybrid controller
matches or exceeds a communication-aware ODrM{+}A* baseline under the
same bandwidth and never underperforms a purely local controller when
bandwidth is scarce; runtime measurements show near-linear scaling in
the number of agents and weak dependence on map size.
Ablations further indicate that imitation learning provides a strong
edge-only baseline, while EEM{+}belief deliver most cloud-side gains by
shifting probability mass away from risky intents.
Overall, the hybrid design approaches centralized performance under
reliable links while retaining distributed scalability under tight
communication budgets, providing a practical route to communication-aware
MAPF at scale.

\newpage
\bibliographystyle{unsrt}
\bibliography{citations.bib}

%HERE Add your bio information along.
% \vspace{-50pt}
\begin{IEEEbiography}[{\includegraphics[width=1in,height=1.25in,clip,keepaspectratio]{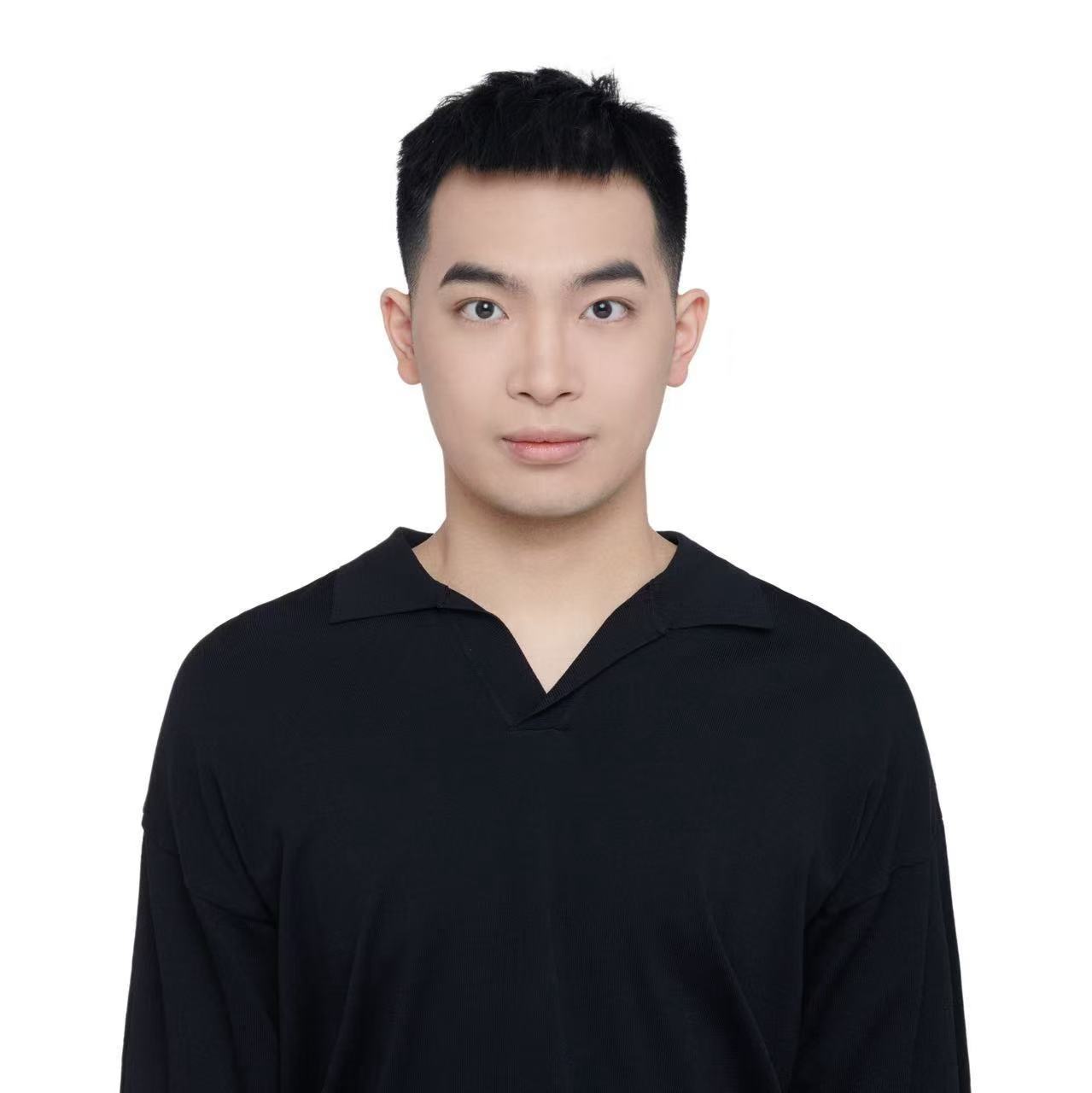}}]{Jinghao Cao} received the BSc in Electrical Engineering Systems and MEng in Electrical and Electronic Engineering from the University of Melbourne, Parkville campus in 2020 and 2022. He is currently a PhD Student at the School of Electrical and Information Engineering, The University of Sydney. His research interests includes wireless network control systems (WNCS), industrial Internet of Things (IIoT), and robotic control systems.
\end{IEEEbiography}

% \vspace{-37pt}
\begin{IEEEbiography}[{\includegraphics[width=1in,height=1.25in,clip,keepaspectratio]{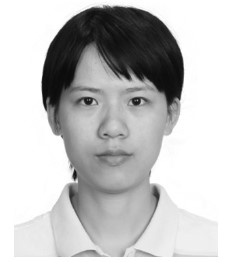}}]{Dr. Wanchun Liu} (Member, IEEE) received the B.S. and M.S.E. degrees in electronics and information engineering from Beihang University, Beijing, China, and Ph.D. from the Australian National University, Canberra, Australia. Following her graduation, she joined the University of Sydney as a research fellow. Her research interest lies in communications and networked control theory, wireless control for industrial Internet of Things (IIoT), and wireless human-machine collaborations for Industry 5.0. She was a co-chair of the Australian Communication Theory Workshop in 2020-2021. She was a recipient of the Australian Research Council’s Discovery Early Career Researcher Award 2023, the Dean’s Award for Outstanding Research of an Early Career Researcher 2022, and the Chinese Government Award for Outstanding Students Abroad 2017.
\end{IEEEbiography}
% \vspace{-35pt}
\begin{IEEEbiography}[{\includegraphics[width=1in,height=1.25in,clip,keepaspectratio]{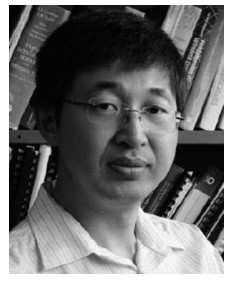}}]{Prof. Yonghui Li} (Fellow, IEEE) received the Ph.D. degree from Beijing University of Aeronautics and Astronautics, Beijing, China, in November 2002.
Since 2003, he has been with the Centre of Excellence in Telecommunications, The University of Sydney, Sydney, NSW, Australia, where he is currently a Professor and the Director of the Wireless Engineering Laboratory, School of Electrical and Information Engineering. His current research interests are in the area of wireless communications, with a particular focus on MIMO, millimeter-wave communications, machine-to-machine communications, coding techniques, and cooperative communications. He holds a number of patents granted and pending in these fields.
Prof. Li is the recipient of the Australian Queen Elizabeth II Fellowship in 2008 and the Australian Future Fellowship in 2012. He received the Best Paper Awards from the IEEE International Conference on Communications 2014, IEEE PIRMC 2017, and IEEE Wireless Days Conferences 2014. He is currently an Editor of the IEEE Transactions on Communications and IEEE Transactions on Vehicular Technology. He also served as the Guest Editor for several IEEE journals, such as IEEE Journal on Selected Areas in Communications, IEEE Communications Magazine, IEEE Internet of Things Journal, and IEEE Access.
\end{IEEEbiography}
\begin{IEEEbiography}[{\includegraphics[width=1in,height=1.25in,clip,keepaspectratio]{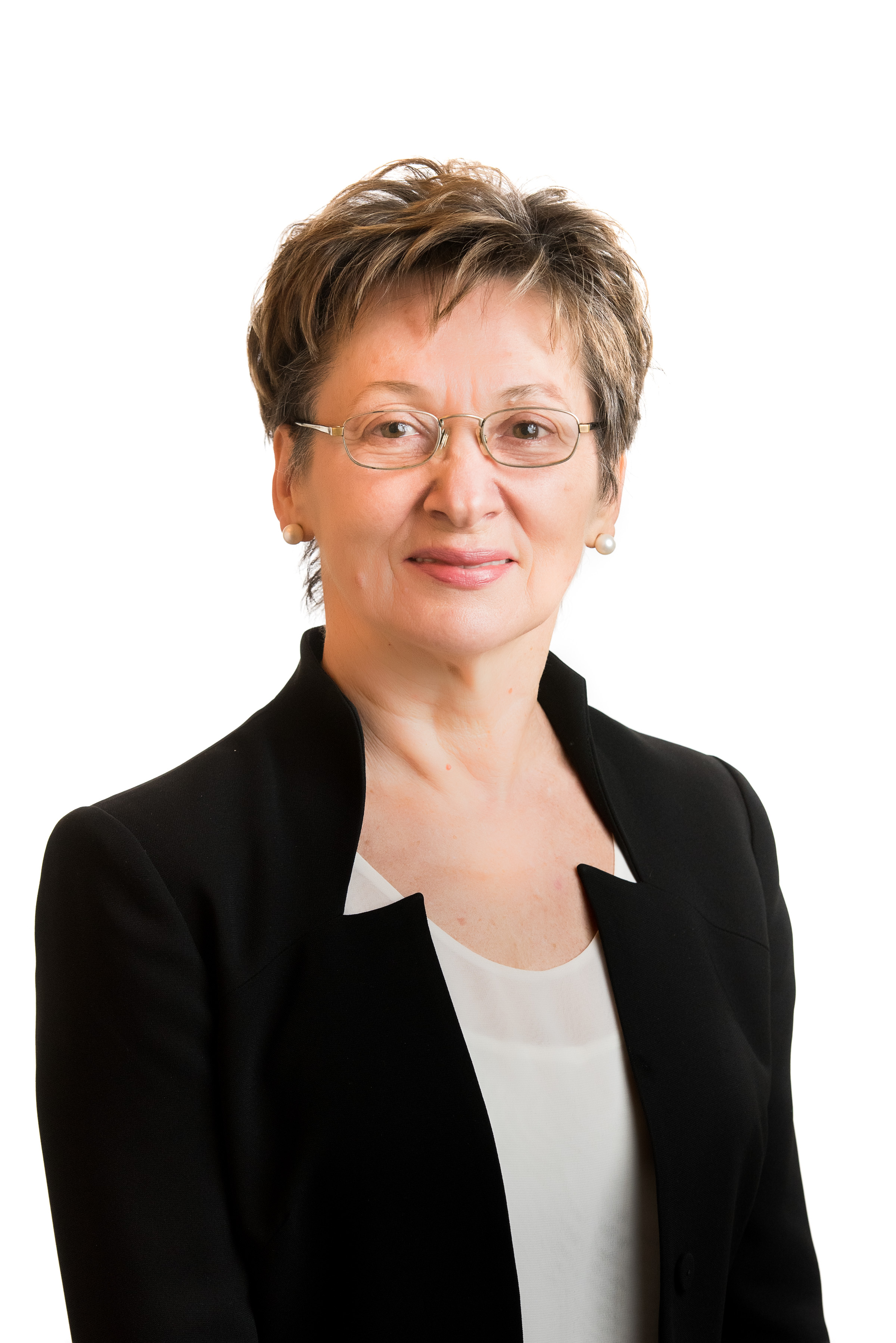}}]{Prof. Branka Vucetic} (Fellow, IEEE)  is an Australian
Laureate Fellow, a Professor of Telecommunications, and Director of the Centre for IoT and
Telecommunications at the University of Sydney.
Her current research work is in wireless networks,
and Industry 5.0. In the area of wireless networks,
she works on communication system design for 6G
and wireless AI. In the area of Industry 5.0, Vucetic’s
research is focused on the design of cyber-physicalhuman systems and wireless networks for applications in healthcare, energy grids, and advanced
manufacturing. Branka Vucetic is a Fellow of IEEE, the Australian Academy
of Technological Sciences and Engineering and the Australian Academy of
Science.
\end{IEEEbiography}
% \vspace{-135pt}

\end{document}